\documentclass[10pt,b5paper,twoside,onecolumn]{scrartcl}
\usepackage[cp1250]{inputenc}
\usepackage{color,fancyvrb}
\usepackage{graphicx}
\usepackage{hyperref}
\usepackage{fancyhdr}
\usepackage{setspace}
\usepackage{float}
\usepackage{pdfpages}
\usepackage{verbatim}
\usepackage{lmodern} 
\usepackage{amsthm}
\usepackage{amssymb}
\usepackage{multicol}
\usepackage{booktabs}
\usepackage{mathpazo}
\usepackage[T1]{fontenc}
\usepackage[titles]{tocloft}
\begin{document}

\pagestyle{fancyplain}
\fancyhf{}
\fancyhead[LE]{\textit{GraphCombEx}}
\fancyhead[RO]{\textit{D. Chalupa, K. A. Hawick}}
\fancyfoot[C]{\thepage}
\fancypagestyle{plain}
{
	\fancyhf{} 
	\renewcommand{\headrulewidth}{0pt} 
	\renewcommand{\footrulewidth}{0pt}
}

\thispagestyle{empty}

\begin{center}\textbf{\LARGE\sffamily\noindent
GraphCombEx: A Software Tool for Exploration of Combinatorial Optimisation Properties of Large Graphs
}\end{center}

\begin{center}{\large\sffamily\noindent David Chalupa$^a$, Ken A. Hawick$^b$}\end{center}
\begin{center}
{
\noindent
$^a$~Operations Research Group\\
Department of Materials and Production\\
Aalborg University\\
Fibigerstr\ae de 16, Aalborg 9220, Denmark\\
Email: \texttt{dc@m-tech.aau.dk}\\
$^b$~School of Engineering and Computer Science\\
University of Hull\\
Cottingham Road\\
Hull HU6 7RX, United Kingdom\\
Email: \texttt{k.a.hawick@hull.ac.uk}

}
\end{center}

\vspace{30pt}

\paragraph{Abstract.} We present a prototype of a software tool for exploration of multiple combinatorial optimisation problems in large real-world and synthetic complex networks. Our tool, called GraphCombEx (an acronym of Graph Combinatorial Explorer), provides a unified framework for scalable computation and presentation of high-quality suboptimal solutions and bounds for a number of widely studied combinatorial optimisation problems. Efficient representation and applicability to large-scale graphs and complex networks are particularly considered in its design. The problems currently supported include maximum clique, graph colouring, maximum independent set, minimum vertex clique covering, minimum dominating set, as well as the longest simple cycle problem. Suboptimal solutions and intervals for optimal objective values are estimated using scalable heuristics. The tool is designed with extensibility in mind, with the view of further problems and both new fast and high-performance heuristics to be added in the future. GraphCombEx has already been successfully used as a support tool in a number of recent research studies using combinatorial optimisation to analyse complex networks, indicating its promise as a research software tool.

\paragraph{Keywords.} GraphCombEx, Graph Combinatorial Explorer, complex net-\linebreak works, large sparse graphs, combinatorial optimisation problems, research software.

\section{Introduction}

The demand for research support software has been growing in the last years in both academic and industrial applications. As an increasing number of real-world problems are concerned with analysis of networked data, exploration of complex networks has become one of the crucial areas of focus, including social networks \cite{visualizationlongitudinalsocialnetworks,optimnetworks}, biological networks \cite{layoutalgorithmbiologicalnetworks} or utility distribution networks \cite{powernetworks,waterdistribution}. This exploration is usually concerned with both numerical and visual analysis of networked data.

From the other perspective, combinatorial optimisation problems and algorithms to solve these problems have been subject of empirical research for a long time. Nowadays, this holds mainly for the variety of well-known NP-hard combinatorial graph-theoretical optimisation problems \cite{karp}. Over the last decade, there have been a vast number of papers on algorithms for solving these problems, ranging from graph colouring \cite{Brelaz,culberson}, through minimum dominating set \cite{chvatal}, to crossing minimisation problems in graph drawing \cite{gadrawingundirected,crossingnphard,esdrawing}. The workflow of experimental research on combinatorial optimisation usually follows a pattern: an experimental algorithm is proposed, independently implemented and tested on a standardised benchmark \cite{JohnsonDimacs}. In some studies, a set of problem instances generated in controlled conditions. The common benchmark usually represents the unifying element.

Even though a number of software tools have been proposed to explore networked data \cite{gephi,csardi2006igraph,graphinvestigator,graphviz,graviz}, this has had little impact on empirical research in combinatorial optimisation so far. In this paper, we present a prototype of a software tool that aims at bridging this gap, providing a unifying platform for exploration of combinatorial properties of large-scale synthetic and real-world complex networks.

\paragraph{Contributions.} The tool is called Graph Combinatorial Explorer (abbr. GraphCombEx). It has been designed and implemented in small steps over the course of several years, to support several studies on combinatorial properties of complex networks. The aim is to provide a unifying platform for efficient exploration of these networks using appropriate scalable algorithms. The core functionality therefore includes a number of constructive heuristics that can be used to quickly find good suboptimal solutions and lower bounds for potentially very large networks. With the NP-hardness of many graph-theoretical problems in mind, the design of GraphCombEx fully acknowledges that scaling and generalisation to previously unseen networks are crucial. However, there are also a few iterative improvement heuristics and the prototype is potentially extensible with more of such algorithms.

GraphCombEx currently supports exploration of undirected graphs with up to $5$ million vertices, which can easily be extended to larger graphs. Graphs are handled in a format similar to the DIMACS format for graph colouring \cite{JohnsonDimacs}. A support tool for conversion from the GML format used in network science is also provided. Several graph generators are used in GraphCombEx, including complete trees, unit disk graphs or scale-free and small-world networks. The tool currently supports suboptimal solution construction and lower bounds for maximum clique and graph colouring, maximum independent set and vertex clique covering, minimum dominating set and the longest simple cycle. A number of graph metrics and visualisation techniques are also supported. Crucially, GraphCombEx is intended as an extensible tool, rather than a one-off project. The aim is to potentially support further problems, algorithms, and metrics in the future. The use of GraphCombEx in several research studies, which have been previously published, is summarised in the use case section of this paper.

The rest of the paper is structured as follows. In Section 2, we review the background and related work from relevant perspectives. In Section 3, we present the overall design elements of GraphCombEx. In Section 4, the summary of evaluation of GraphCombEx in several previous projects is presented, along with a discussion. Last but not least, Section 5 summarises this work and hints the future research directions.

\section{Background and Related Work}

At the start of this section, it is first worth noting the reasons for designing a new software platform for these types of problems. In fact, GraphCombEx draws inspiration from a number of research projects described below. However, combinatorial optimisation problems in complex networks tend to have several very specific properties that need to be addressed in software design. These can range from special data structures to complex parameterisations. Each problem and each algorithm tend to be quite specific.

If exploring a previously unseen network, not only that the optimum for many combinatorial properties is not known, it is often not known how hard it will be to find. It is possible that a simple heuristic will be able to find a proven optimum. However, it is also possible that a highly advanced algorithm is required. In that case, such a requirement should first be confirmed by the use of a simpler heuristic. In addition, the network to study may potentially be huge. For small but structurally intricate networks, one can use high-performance algorithms to discover previously unknown problem solutions. For large networks, it may be non-trivial to discover any reasonable solution. This has an impact on the graph representation, selection of the right heuristics that scale well in terms of computational time, as well as the choice of the right data structures. These observations have a significant impact on the overall design of a tool that should be useful for practice. Such specific demands are perhaps also the reason why combinatorial optimisation problems in complex networks are rarely explored by out-of-the-box solutions.

Several software packages for complex network exploration and visualisation have been proposed and developed over the last years. We will now provide a brief overview of some of these solutions.

The igraph library is one of the most popular open-source software packages for network exploration \cite{csardi2006igraph}. It is available in several variants, including interfaces for C, Python and R. The package provides a general-purpose platform for handling of both undirected and directed graphs. Solving routines for several classical graph-theoretical and graph mining problems are supported, e.g. maximum flow, minimum spanning tree or cluster detection.

Another general-purpose solution is the SNAP network analysis and graph mining library \cite{leskovec2016snap}. This software package is intended as a high-performance large-scale complex network exploration tool, with support for handling potentially huge graphs. This platform is focused primarily on very large-scale network analytics, using a variety of metrics and simulations on the networks.

Graph drawing and visualisation is another significant aspect of complex network exploration software \cite{graphdrawing,TamassiaHandbookOfGraphDrawing}. One of the most successful open-source software platforms focused mainly on graph visualisation is Gephi \cite{gephi}. The features supported by Gephi include network exploration, interactive manipulation, 3D rendering, as well as of image exporting.

Another popular software package for graph visualisation is GraphViz \cite{graphviz}. This package is designed as a collection of customisable graph drawing tools, including batch layout and graph editors.

GraViz prototype \cite{graviz} is aimed at interactive visualization and exploration of graphs. This tool is particularly well-suited for geographical data. It has been used to explore water distribution networks \cite{waterdistribution} or power networks \cite{powernetworks}, for which it is easy to characterize their vertices by coordinates.

Another specific property of NP-hard combinatorial optimisation problems in networks is that these are often formulated as integer linear programs. Out-of-the-box mixed integer linear programming (MILP) software tools are increasingly used in solving these problems in practical applications \cite{BonamiAlgorithmicMixedIntegerPrograms,LinderothMilp}. However, open-source MILP software and network exploration software still seem to be perceived in relative isolation, which is another interesting area of challenge.

\section{GraphCombEx: Graph Combinatorial Explorer and its Design}

The general idea behind GraphCombEx is quite simple and emerged from a number of similar problem solving projects. Usually, if several combinatorial optimisation problems and algorithms are explored, a suite of isolated ad-hoc tools are developed, especially if new algorithms are designed. In this research, a single unified solution for several problems and algorithms was attempted as an alternative.

In principle, GraphCombEx is designed as a platform for analysis and experimentation with static large sparse graphs. Specific focus is put on large-scale complex networks, as these have been gaining increasing attention in the last years. With particular interest, GraphCombEx addresses NP-hard combinatorial optimisation problems. Its design therefore reflects the need to incorporates several efficient and well scalable heuristics to compute lower and upper bounds and high-quality suboptimal solutions to these problems for potentially very large complex networks.

GraphCombEx currently supports undirected graphs with up to $5$ million vertices. This bound was only chosen for better predictability in memory management, computational complexity and testability, even though this bound can in theory be extended by changing a single constant in the source code. The graphs are loaded and saved into an extension of the COL format known in DIMACS graph colouring benchmarks \cite{JohnsonDimacs}. This format has been chosen to achieve simplicity. An example of a simple social network in this format is given in Listing 1. The tree-based visualisation and the adjacency matrix visualisation, generated by GraphCombEx for this particular network, are given in Figure 1.

It is worth noting that a simple tool is also available with GraphCombEx, to allow conversion of the popular GML format, used in network science, to the extended COL format.

\begin{figure}
\begin{center}
Listing 1. Representation of a simple artificial social network graph in the extended DIMACS graph format.
\noindent\rule{\textwidth}{1pt}
\end{center}
\begin{verbatim}
c 1 Alice
c 2 Bob
c 3 Cindy
c 4 Daniel
c 5 Emily
c 6 Frank
p edge 6 8
e 1 2
e 1 3
e 1 5
e 2 3
e 2 4
e 2 6
e 3 4
e 4 5
\end{verbatim}
\noindent\rule{\textwidth}{1pt}
\end{figure}

\begin{figure}
\begin{center}
\includegraphics[scale=0.3]{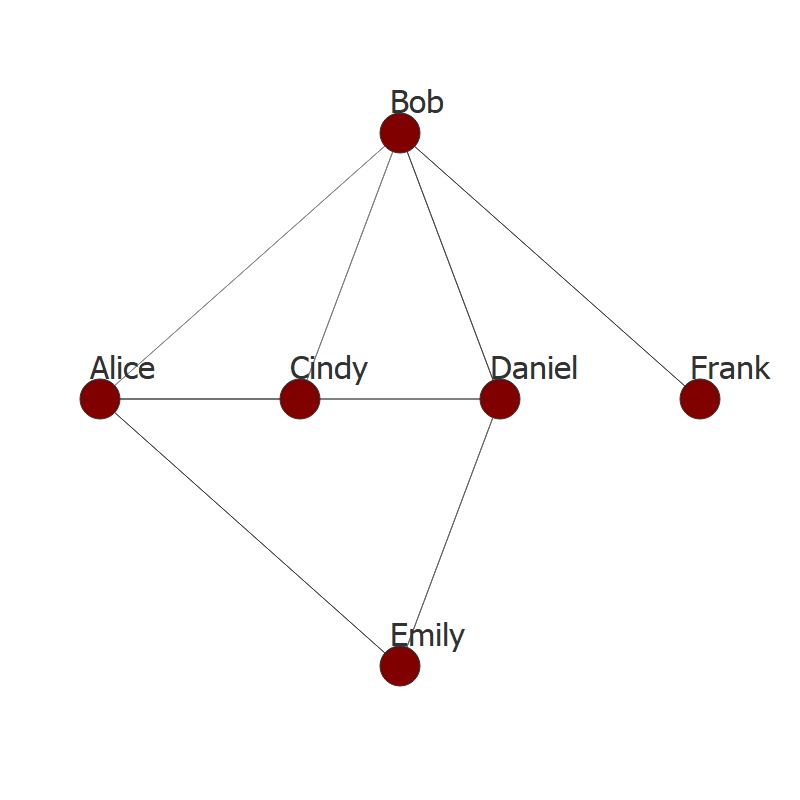}\hspace{20pt}
\includegraphics[scale=38]{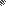}
\end{center}
\caption{A tree-based drawing and an adjacency matrix visualisation of the artificial social network defined in Listing 1, as generated by GraphCombEx.}
\end{figure}

GraphCombEx supports generating multiple types of types of graphs. These currently include:

\begin{itemize}
	\item complete $n$-ary trees;
	\item unit disk graphs;
	\item scale-free networks generated by the Barab\'asi-Albert preferential attachment model \cite{barabasialbert,barabasialbert2};
	\item regular grids with probabilistic rewiring;
	\item small-world networks generated by the Watts-Strogatz model \cite{WattsStrogatzCollectiveDynamicsSmallWorldNetworks}.
\end{itemize}

\noindent
Several simple graph manipulation and preprocessing routines are also included:

\begin{itemize}
	\item generation of complementary graphs;
	\item pruning of leaves (iterative, i.e. the resulting graph will contain only vertices with degree at least $2$);
	\item isolating the largest connected component;
	\item generation of shortcut graphs used in solving $k$-reachability problems \cite{CHALUPA20171}.
\end{itemize}

\noindent
GraphCombEx computes a number of metrics for the graph explored. Those, which are currently supported, include:

\begin{itemize}
	\item the numbers of vertices, edges, connected components, density, minimum, maximum, average and standard deviation of degrees;
	\item the number of triangles and the mean clustering coefficient of a vertex;
	\item graph girth (length of the shortest cycle) and maximum diameter of an isolated connected component.
\end{itemize}

\noindent
Some of these are computed upon loading or generating the graph, as long as the complexity of the corresponding algorithms is at most $\mathcal{O}(m)$, where $m$ is the number of edges. The numbers of vertices, edges, minimum, maximum and average degree, standard deviation of the degree are computed directly when graph is loaded or generated. Upon request, the number of triangles, the mean clustering coefficient, and the diameter can be computed. The number of triangles is computed using scanning triplets of adjacent vertices and may potentially take a long time for very large graphs. The same holds also for the mean clustering coeffient of a vertex. The maximum diameter of a connected component is computed in $\mathcal{O}(nm)$ time, where $n$ is the number of vertices.

Within GraphCombEx, the graph is represented as adjacency lists with sorted adjacencies, to allow for binary search to be used for efficient checking of adjacencies. An alternative extension would include a use of a hash table for direct adjacency checking. This would speed up some computationally intenstive routines but is not supported in the pilot version, as lower memory demands were preferred.

The main part of functionality of GraphCombEx consists of several heuristics and approximation algorithms, which can be used to compute bounds and suboptimal solutions to NP-hard graph-theoretical problems. The currently supported problems include:

\begin{itemize}
	\item maximum clique and graph colouring / chromatic number problems (representing reciprocal lower and upper bounds);
	\item maximum independent set and vertex clique covering problems (representing reciprocal lower and upper bounds);
	\item minimum dominating set problem (both a suboptimum and simple lower bounds);
	\item the longest cycle problem (both a suboptimum and simple upper bounds).
\end{itemize}

\noindent
These problems have been chosen for their high relevance for practice, as well as for their relatively diverse representations and properties. Maximum clique, maximum independent set and minimum dominating set problems represent 0-1 constrained substructure detection problems. Graph colouring and vertex clique covering are representatives of vertex labelling problems. Last but not least, the longest cycle problem is a problem of identification of specific walks on network. Extensions to other 0-1 optimisation problems (e.g. vertex cover), walk problems (e.g. longest path) or labelling problems (e.g. various community detection problems) should therefore be possible in the chosen framework.

The core of GraphCombEx is represented by techniques for computing bou\-nds and high-quality suboptimal solutions to combinatorial problems in large sparse graphs. The heuristics currently supported include both fast constructive algorithms, as well as a few more computationally intensive iterative improvement algorithms. The constructive algorithms include the following heuristics and approximation algorithms:

\begin{itemize}
\item For maximum clique, GraphCombEx uses a greedy heuristic with binary heap, which orders the vertices by degree from largest to smallest, similar to the greedy heuristic for maximum independent set \cite{gis}.
\item For the chromatic number problem, Br\'{e}laz's heuristic DSATUR is employed \cite{Brelaz}. It implementation with binary heap is used, to allow for quick computing of an upper bound of the chromatic number \cite{saturation,easycolor}
\item Maximum independent set is estimated by the greedy approximation heuristic, which orders the vertices by degree from smallest to largest. In our implementation, we use a binary heap. This heuristic guarantees $\mathcal{O}(\Delta)$-appproximation, where $\Delta$ is the maximum degree of a vertex in the graph \cite{gis}.
\item Minimum dominating set is approximated by a classical greedy algorithm, originally proposed for the set covering problem. This algorithm guarantees an $\mathcal{O}(\log(\Delta))$-appproximation for an arbitrary graph \cite{chvatal}.
\item The longest cycle problem is solved by a heuristic based on depth-first search or its improvement with local search \cite{CHALUPA201796}.
\end{itemize}

\noindent
The iterative improvement algorithms are used as longer-running procedures that improve an initially generated solution. These are therefore executed in a separate thread and their intermediate results are updated within the user interface over the course of the computation:

\begin{itemize}
	\item For the chromatic number and maximum clique problems, GraphCombEx uses the iterated greedy (IG) graph colouring algorithm \cite{culbersontechreport,culberson}, enhanced by an order-based randomized local search (RLS) for maximum clique, similar to RLS used for maximum independent set.
	\item For combined maximum independent set and minimum vertex clique covering, GraphCombEx employs the IG clique covering algorithm, combined with RLS for maximum independent set \cite{cai15}.
\end{itemize}

\noindent
For graphs with bounded numbers of vertices and edges, the GraphCombEx prototype also supports several different forms of their visualization:

\begin{itemize}
	\item centrality-based (also sometimes referred to as radial);
	\item grid-based;
	\item tree-based (simple hierachical visualisation);
	\item circular.
\end{itemize}

\noindent
Additionally, GraphCombEx supports visualisation of the currently longest cycle found, if the heuristic is used. The vertices of this cycle can also be rearranged as an ``outer circle'', in a modification of the centrality-based visualisation. This is referred to as a cycle-based visualisation and was used in the study on the longest cycle problem \cite{CHALUPA201796}.

In the centrality-based (radial) visualization, the vertex with maximum degree is put in the middle and other vertices are placed in levels, based on the distance to the centre, similarly to the GraViz prototype \cite{graviz}. Radial drawings are quite popular in graph visualisation in general \cite{radialsugiyama,radiallayout}. In the grid-based visualization, the vertices are distributed simply in a regular grid. The third visualisation is the tree-based visualisation, in which the vertex with maximum degree is placed at the top and the other vertices are arranged hierarchically, based on the distance to this vertex. This was chosen as a rather simple implementation of one of the hierachical drawings methods \cite{drawingtrees,hierarchicalgraphvisualization}. The last drawing method supported is the circular drawing method \cite{crossingreductioncircular}.

Our prototype also visualizes a bitmap, which represents the adjacency matrix of the graph. Visualization of the shape of degree distribution is used without restrictions on the size of the graph. Export features for graph visualisation, as well as adjacency matrix visualisation and degree distribution in CSV format are provided.

\begin{figure}
\begin{center}
\includegraphics[scale=0.2]{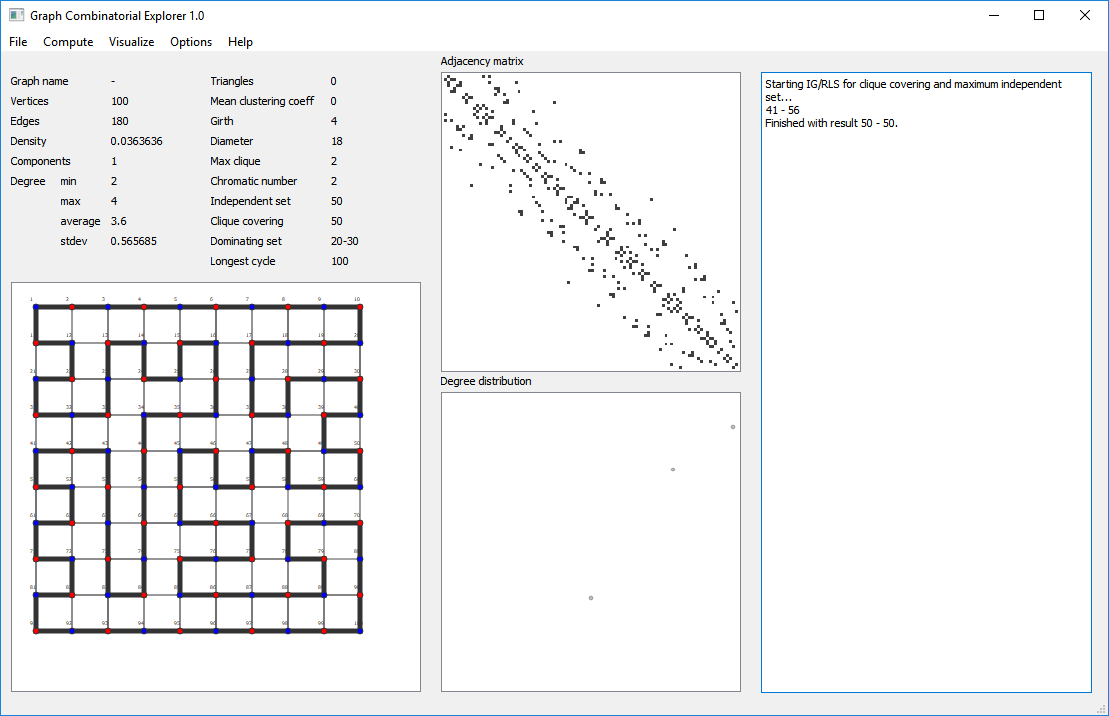}
\includegraphics[scale=0.2]{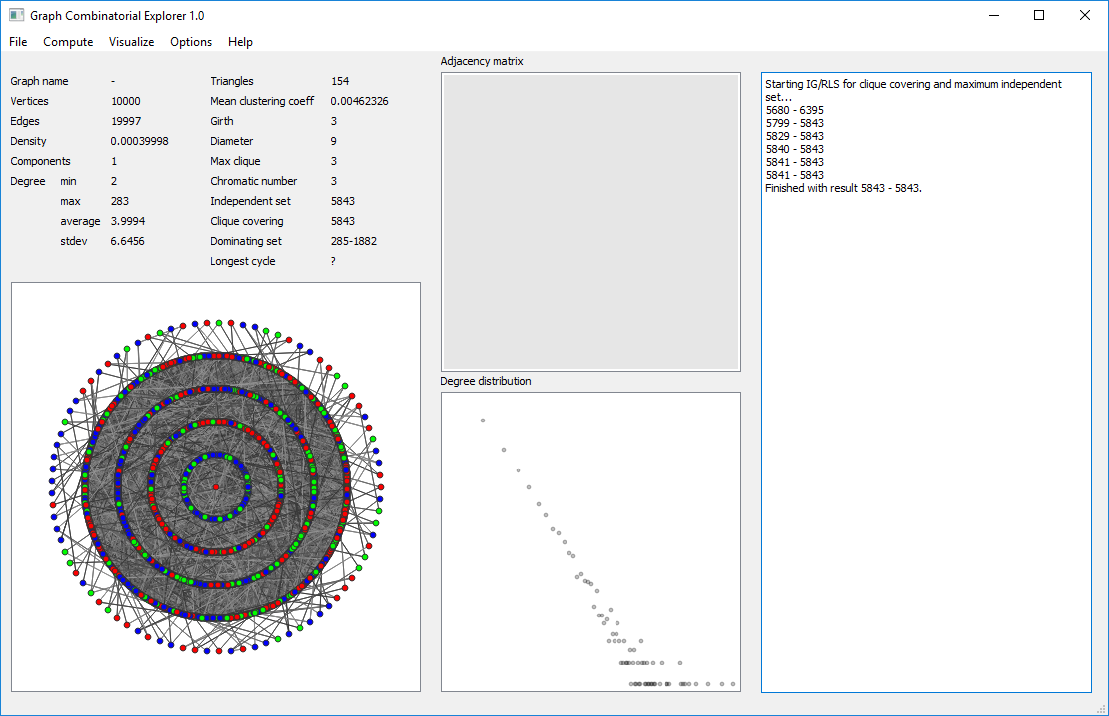}\\
(a) \hspace{160pt} (b)\\
\includegraphics[scale=0.2]{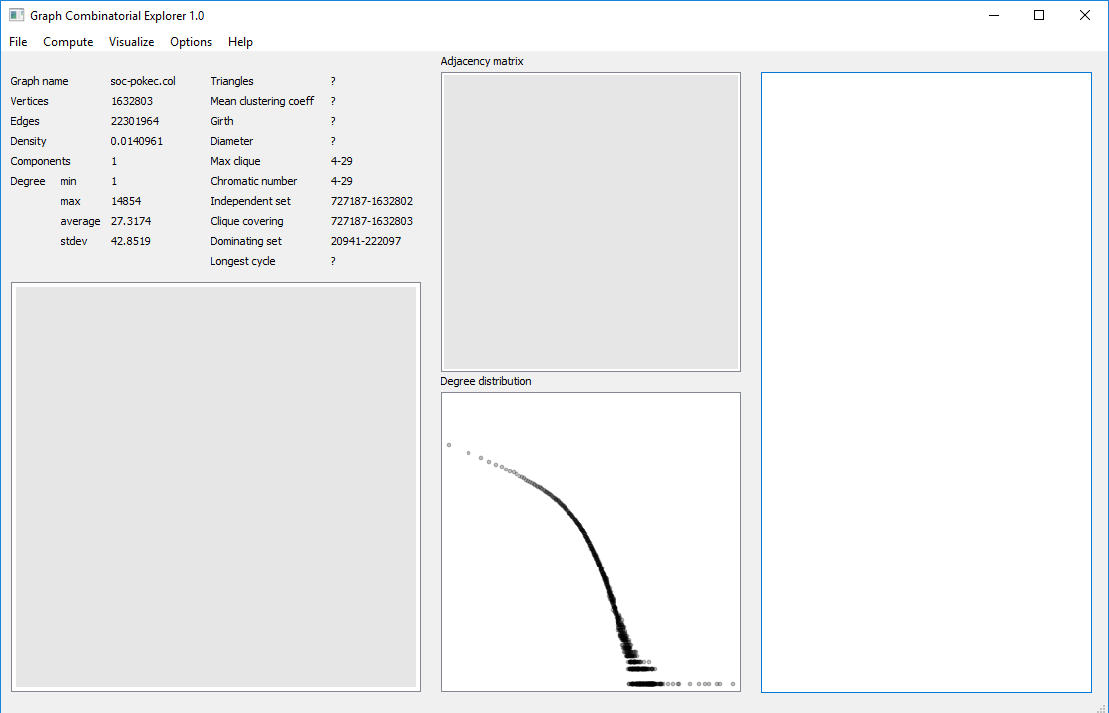}
\includegraphics[scale=0.2]{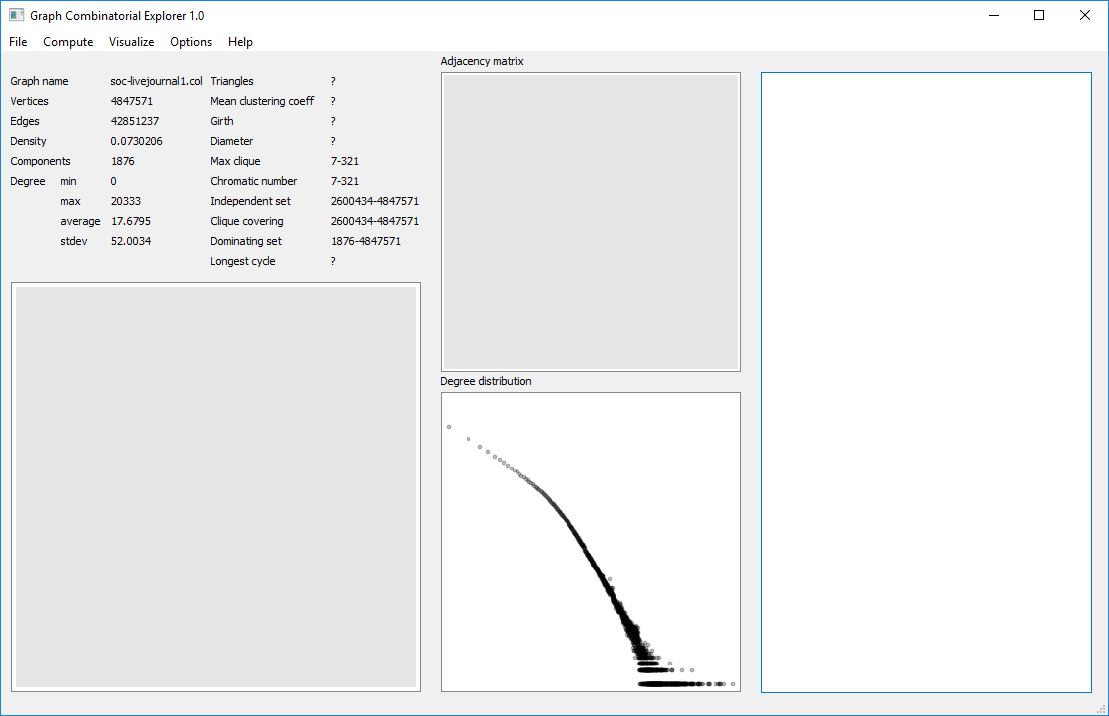}\\
(c) \hspace{160pt} (d)\\
\end{center}
\caption{Graphical user interface (GUI) of GraphCombEx prototype, exploring four different complex networks. Network (a) is a $10 \times 10$ regular grid, while network (b) is a scale-free network on $10000$ vertices generated by Barab\'{a}si-Albert model. Networks (c) and (d) are large-scale social snapshots of networks Pokec and LiveJournal from the SNAP network data repository. GUI shows a visualisation of the graph and its adjacency matrix, if the graph size is within a suitable range for the visualisation. Degree distribution shape is also depicted and properties are computed for each graph, including ranges for the optima of the NP-hard problems studied.}
\end{figure}

GraphCombEx was implemented in C++ using the Qt framework and is available under the GNU General Public License v3. It can therefore be used on several operating systems, including Windows, as well as Unix-like systems. General information on GraphCombEx is freely available online\footnote{\url{http://davidchalupa.github.io/research/software/graphcombex.html}}, along with the source code repository on GitHub\footnote{\url{https://github.com/davidchalupa/GraphCombEx}}.

\section{Use Cases for GraphCombEx and Discussion}

Evaluating software tools and prototypes with a suitable level of rigour is a complex task, especially when the software is designed to explore complex systems such as the networks studied here. We first illustrate a typical scenario of the use of GraphCombEx. Next, we will summarise a number of studies, in which GraphCombEx has already been successfully used to support the research projects. We hope that GraphCombEx will further be extended by more algorithms, problems to explore and used in further projects. This followed by a discussion to provide further context, as well as the way forward in extensions and applications of this prototype.

Figure 2 shows the main graphical user interface, illustrating the main features for four complex networks. Networks explored in (a)-(b) are a regular grid and a scale-free network that were generated, while (c)-(d) represent large-scale social networks from the SNAP network data repository \cite{LeskovecSNAP}. These illustrate the functionality in an overview, including the metrics provided, as well as the problems and analytics supported.

If GraphCombEx is built into a 64-bit binary, it can be used to explore relatively large graphs. Networks (c)-(d) within Figure 2 are large snapshots of social networks Pokec \cite{pokecanalysis} and LiveJournal \cite{BackstromGroup,LeskovecCommunity}. The latter consists of more than $4.8$ million vertices and a 64-bit version of GraphCombEx is required to process a network of this size.

One could argue that potential applications of GraphCombEx are numerous, which stems from its relatively general design. The combinatorial properties supported and heuristics provided indeed represent a very general view and are specifically beneficial for initial exploration of previously unseen networked data. However, specialised computational problems related to those supported by GraphCombEx range from scheduling and routing problems in operations research, through data analytics for social media \cite{visualizationlongitudinalsocialnetworks,optimnetworks}, to interaction exploration in bioinformatics in support of drug discovery \cite{molecularnetworksdrugdiscovery}. This opens a number of pathways for applications of the tool.

\paragraph{A typical use case for GraphCombEx.} We will illustrate the power of GraphCombEx using the simple small network presented in Listing 1. In principle, this use case can also be applied to larger networks, even though the tractability of more computationally intensive routines may depend on the size of the graph. After opening the graph file using \textit{File $>$ Open}, one will see the centrality-based visualisation by default. By using \textit{File $>$ Visualization type}, the graph visualization mode can be changed. By using the options within the \textit{Compute} menu, one can compute the on-demand characteristics of the network. For the network from Listing 1, the values presented in Table 1 will be obtained. GraphCombEx finds the two triangles in the networks and computes the mean clustering coefficient. Maximum clique size is $3$ and the graph is $3$-colourable. Maximum independent set and clique covering of size $3$ are also computed. Dominating set of size $2$ is also found. From the drawings, one can see that this dominating set is $\{Bob,Emily\}$. The longest cycle of size $5$ is also found. All these values are proven optima for the network. If only bounds were computed, GraphCombEx would present intervals for each optimal value. An illustration of the $3$-colouring and the longest cycle on $5$ vertices in different graph layouts supported by GraphCombEx is given in Figure 3.

\begin{table*}
{\scriptsize
\begin{center}
\caption{The values of metrics and bounds computed by GraphCombEx for the simple social network from Listing 1.}
\begin{tabular}{l l}
\toprule
metric / bound & value \\
\midrule
triangles & 2 \\
mean clustering coefficient & 0.277778 \\
girth & 3 \\
diameter & 3 \\
maximum clique & 3 \\
chromatic number & 3 \\
independent set & 3 \\
clique covering & 3 \\
dominating set & 2 \\
longest cycle & 5 \\
\bottomrule
\end{tabular}
\end{center}}
\end{table*}

\paragraph{Support for exploration of very large sparse graphs.}
GraphCombEx has been employed to support several previous studies on combinatorial optimisation in large sparse graphs. It has been partially used to support the previous research on detection of long cycles in real-world complex networks \cite{CHALUPA201796}. This included social network samples, as well as protein-protein interactions. In addition, it supports generation of shortcut graphs used in solving the $k$-reachability problem, in which one aims to find the smallest set such that each vertex is in distance at most $k$ to at least one vertex of such a set \cite{CHALUPA20171}. The iterated greedy heuristic for vertex clique covering is also supported in GraphCombEx \cite{ist14,cai15}. A typical application area for these studies is analysis of large-scale social networks.

\paragraph{Analysis of scientific collaboration networks.}
Collaboration networks are of a high interest within the scientific community \cite{NewmanScientificCollaborationNetworks}. Combinatorial properties can offer further insights into the phenomena of scientific collaboration patterns and their understanding. In particular, the $k$-reachability problem has already been studied for several scientific collaboration networks, with partial support of GraphCombEx. It has been revealed that for most collaboration networks studied, there is a single vertex that is roughly in a distance between $8$ to $11$ to other vertices of the collaboration networks \cite{CHALUPA20171}.

\begin{figure}
\begin{center}
\includegraphics[scale=0.24]{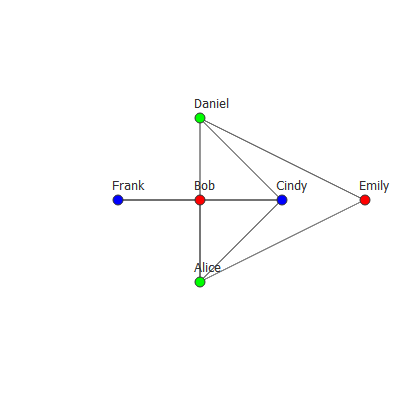}\hspace{35pt}
\includegraphics[scale=0.24]{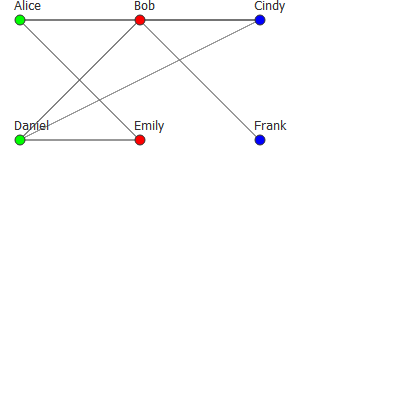}\hspace{0pt}
\includegraphics[scale=0.24]{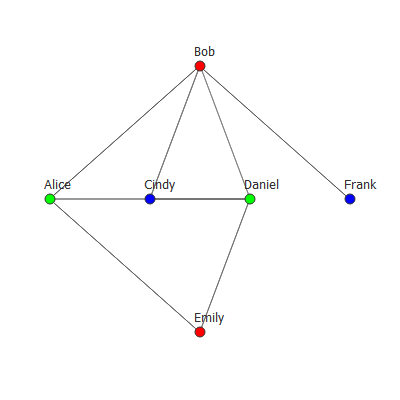}\hspace{10pt}
\includegraphics[scale=0.24]{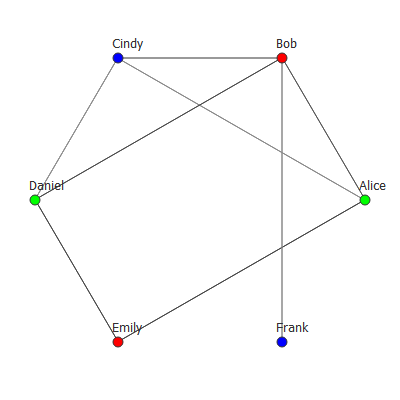}\hspace{10pt}
\includegraphics[scale=0.24]{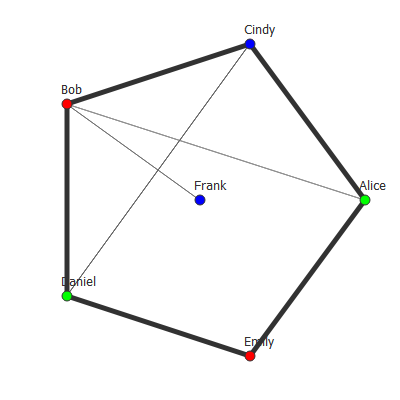}
\end{center}
\caption{An illustration of the visualisation types provided by GraphCombEx, applied to the simple social network from Listing 1. GraphCombEx supports different layouts, as well as highlighting of the colouring and the longest cycle identified.}
\end{figure}

\paragraph{Exploration of protein-protein interactions.}
One application area of a high interest for GraphCombEx is the study of protein-protein interaction networks. A general study of protein-protein interactions and combinatorial optimisation has been conducted with the use of GraphCombEx \cite{evobio16}. Further studies can be conducted, especially given the rich data currently available, e.g. at the UCLA database of interacting proteins \cite{SalwinskiDatabaseOfInteractingProteins,XenariosDatabaseOfInteractingProteins,XenariosDatabaseOfInteractingProteins2001,XenariosDatabaseOfInteractingProteins2002}. This data on protein-protein interactions has also been tested and used within the framework of GraphCombEx \cite{CHALUPA201796}.

\paragraph{Modelling of randomised lattice-based systems and utility distribution networks.}
One of the interesting applications of GraphCombEx, that are still left to be largely explored, is investigation of lattice-based network models. These potentially have structures similar to power networks \cite{powernetworks} or water distribution networks \cite{waterdistribution}. Effects of edge rewiring, insertion or deletion on long cycles, flows and robustness of these networks can be explored using efficient mechanisms, expanding on the existing studies \cite{CHALUPA20171,CHALUPA201796}. Adding the spatial information for these networks is another interesting future direction.

This list is not necessarily exhaustive and highlights that the potential of GraphCombEx as a multi-perspective complex network analysis and visualisation tool. GraphCombEx has been demonstrated as useful in analysis of social network samples of different scales or analysis and visualisation of protein-protein interactions. Other applications can follow, including utility distribution networks, further exploration of biological networks such as gene expression data, or scheduling and transportation networks in operations research. In addition, there is further space for extending GraphCombEx with new combinatorial properties and new algorithms, including local search and evolutionary algorithms or swarm intelligence algorithms approaches.

Furthermore, these algorithms and applications can further be facilitated by integration with integer linear programming (ILP) problem solving tools. As the first example, GraphCombEx currently provides the functionality to export ILP-based representation of the minimum dominating set problem, along with its LP relaxation. MPS linear programming instance format is currently used for this export. MILP software can be used to solve these instances, e.g. the CBC branch-and-cut solver from the COIN-OR package \cite{BonamiAlgorithmicMixedIntegerPrograms,LinderothMilp}.

\section{Conclusions}

We presented a prototype of Graph Combinatorial Explorer (abbr. GraphCombEx), which is an open-source software platform aimed at unified exploration of combinatorial properties of large networks.
GraphCombEx takes the specific properties of the numerous NP-hard combinatorial optimisation problems in networks into account in its design.
The tool currently supports graphs with up to $5$ million vertices, even though this bound is extensible and has been chosen for practical reasons such as memory management.

GraphCombEx employs several heuristic procedures to find high-quality solutions to NP-hard graph problems. These problems currently include maximum clique, chromatic number, maximum independent set, minimum vertex clique covering, minimum dominating set, as well as the longest simple cycle problem. However, the tool is designed with extensibility by further problems and algorithms in mind, including iterative improvement algorithms. The core functionality consists mainly of constructive heuristics enhanced by efficient data structures, that can be used to find good suboptimal solutions and bounds for very large graphs.

Complex networks with no explicit spatial properties are particularly addressed in the design of the prototype of this software tool. GraphCombEx has been successfully used as a support tool in a number of studies, including exploration of combinatorial properties of protein-protein interactions \cite{evobio16}, detection of long simple cycles \cite{CHALUPA201796}, or the $k$-reachability problem in small-world complex networks \cite{CHALUPA20171}, including social networks or scientific collaboration networks.

Possible future applications of GraphCombEx include analysis of massive public social networks \cite{pokecanalysis}, further exploration of protein-protein interaction networks \cite{proteincentrality}, metabolic pathways \cite{metabolicpathwaylayout}, or utility distribution networks \cite{powernetworks,waterdistribution}.

\bibliography{common}{}
\bibliographystyle{plain}

\end{document}